\documentstyle[11pt,newpasp,twoside,psfig]{article}
\begin{document}
\title{Learning about Jets from Observations of Blazars}
\author{Marek Sikora}
\affil{N. Copernicus Astronomical Center, Bartycka 18, 00-716 Warsaw, Poland}
\author{Greg M. Madejski}
\affil{Stanford Linear Accelerator Center, Menlo Park, CA 94025, USA}

\setcounter{page}{1}
\index{Sikora, M.}
\index{Madejski, G.}

\begin{abstract}
Jets, paving their way outward through the inner regions of active nuclei, 
Compton-interact with the UV radiation from an accretion disc and broad 
emission line region.  We calculate the predicted properties 
of the resulting spectral signatures of this bulk-Compton process, 
noting that they  are {\sl independent} on the fractional proton 
content or kinetic power of the jet, and use the presence 
or absence of such signatures to put constraints on the 
structure of jets near their bases. 
\end{abstract}

\section{Introduction}

As it was pointed out by Begelman \& Sikora (1987), the bulk-Compton 
interaction of a jet with the UV radiation of an accretion disc  
should lead to the production of the soft X-ray ``bump'' 
in the spectra of FSRQ 
(flat-spectrum-radio-quasars).  Furthermore, the velocity modulation 
of the jet flow by the central engine should result in a formation of 
soft X-ray precursors of non-thermal flares produced in internal 
shocks (Sikora \& Madejski 2002).  The strength of such soft X-ray 
spectral features depends on the bulk Lorentz factor of a jet, $\Gamma$,
the electron number flux, $\dot N_e$, and the energy density of the ambient 
radiation field, $u_{ext}$. Since $u_{ext}$ depends on the distance
from the central engine, constraints imposed by observations on 
the magnitude of the bulk-Compton features can be used to probe the 
spatial scale of the jet formation process.  We calculate the luminosity 
of the soft X-ray  features using $\Gamma$ and $\dot N_e$ as derived 
from the EC (external Compton) model of $\gamma$-ray production in 
FSRQ (Sikora, Begelman, \& Rees 1994) for two jet models:  in one, 
the flow is steady-state, and in another, the jet is composed from 
discrete ejecta.  In the former case, non-thermal events can be 
powered by reconnection of magnetic fields, as it is likely to take place 
in the Poynting flux dominated flows (see, e.g. Drenkhahn \& Spruit 2002; 
Blandford 2000);  in the latter case they are powered by 
collisions of ejecta which lead to formation of internal shocks 
(Sikora et al. 1994, 2001; Spada et al. 2001).
We demonstrate that the lack of prominent soft X-ray excesses, 
and of soft X-ray precursors of the non-thermal flares in available data 
implies that quasar jets are formed on scales $> 10^{17}$cm, and/or 
that inner parts of an accretion disc are radiatively inefficient, 
where the angular momentum is transported outwards by other means 
than viscosity in the disk.  


\section {Non-thermal radiation}
One of the biggest surprises inferred from the EGRET data was the finding 
that during the high states of FSRQ the $\gamma$-ray fluxes exceed 
those in other spectral bands by a very large factor, $\sim$ 10-100
(von Montigny et al. 1995).  Soon after that discovery it was realized that 
the exceptionally high  $\gamma$-ray luminosities of FSRQ can result from 
Comptonization of external radiation fields.  Indeed, the data collected 
during the entire period of the CGRO mission strongly supports that idea.  
All main features of the high energy spectra of FSRQs during outbursts 
can naturally be explained in terms of the external-Compton (EC) model 
(see review by Sikora \& Madejski 2001).  In particular, the distances of
production of short $\gamma$-ray flares in a jet, as inferred from their
variability time scales, agree well with the estimates of the distance 
based on the assumption that the spectral break observed in the 
1 -- 30 MeV range results from the cooling break in the electron energy
distribution. 

In the simplest version of the EC model, electrons responsible for 
the non-thermal radiation in FSRQ are injected with a single power law 
distribution, $Q = K \gamma^{-p}$. They cool and evolve into a 
double-power law distribution, $N_{\gamma} \propto \gamma^{-s_{l,h}}$, 
with $s_h = p+1$ for $\gamma > \gamma_c$, and with $s_l = p$ for  
$\gamma < \gamma_c$, where the $\gamma_c$ is determined by the 
equality of time scales of injection and electron energy loss.  
Electrons with $\gamma >\gamma_c$ produce $\gamma$-ray radiation with 
the energy flux distribution $F_{\nu} \propto \nu^{-\alpha_{\gamma}}$, 
where $\alpha_{\gamma} = (s_h -1)/2 = p/2$, and electrons with 
$\gamma <\gamma_c$ produce X-rays  with  $\alpha_x = (s_l-1)/2= (p-1)/2$.  
Hence, the slope of the injected electrons, $p$, can be recovered from 
the slope of X-ray or $\gamma$-ray spectrum.  Since the soft/mid X-rays 
are likely to be diluted by the contribution from the synchrotron-self Compton
process (Kubo et al. 1998), a more reliable approach is to 
infer $p$ from the $\gamma$-ray spectra.  For FSRQ, $\alpha_{\gamma} \sim 1$ 
(Pohl et al. 1997), and therefore $p = 2$ can be used as a fiducial index.  

Normalization of the electron injection function, $K$,  is also based
on the $\gamma$-ray data. Equating the $\gamma$-ray luminosity 
at a given frequency with the electron emissivity and
noting that in the fast cooling regime ($\gamma > \gamma_c$)
$N_{\gamma} \vert d\gamma/dt' \vert \simeq \int_{\gamma} Q d\gamma$,   
we obtain 
$$ K \simeq 
{2 (\nu_{\gamma} L_{\nu_{\gamma}}) \over  m_e c^2} \,
{\Gamma^2 \over {\cal D}^6}
\, , \eqno(1) $$
where ${\cal D} = [\Gamma(1 - \beta_{\Gamma}\cos{\theta_{obs}})]^{-1}$.
(Note that the factor $\Gamma^2/{\cal D}^6$ comes from the fact that 
for the EC process  $L \simeq ({\cal D}^6/\Gamma^2) L'$ [Dermer 1995]).   
Hence, the total number of relativistic electrons involved in production 
of a nonthermal flare is 
$$ N_e \simeq {\cal D}t_{fl} \int_1 Q \, d\gamma \simeq {\cal D}t_{fl} K
\simeq {2 t_{fl} (\nu_{\gamma} L_{\nu_{\gamma}}) \over m_e c^2} \,
{\Gamma^2 \over {\cal D}^5}
\, , \eqno(2) $$
where $t_{fl}$ is the observed time scale of the $\gamma$-ray flare. 

\section{Bulk-Compton radiation}
{\it Steady-state flow}

Electrons which prior to the dissipative event are cold and
are streaming  steadily through the external UV radiation field
produce soft X-ray radiation with the apparent luminosity
$$ L_{BC} \simeq {\cal D}^2 \int n_e c\sigma_T u_{diff} \Gamma^2 \, dV 
\, , \eqno(3) $$
where $dV= \Sigma dr$ is the volume element and $n_e$ is the electron number
density. Noting that $n_e \simeq \dot N_e /( c \Sigma)$, where 
$\dot N_e \sim N_e/(\lambda/c)$ is the flux of electrons and 
$\lambda \simeq ct_{fl}$ 
is the longitudinal extension of the non-thermal source, we obtain
$$ L_{BC} \simeq {2 \sigma_T \over m_e c^2} {\Gamma^4 \over {\cal D}^3}
\, (\nu_{\gamma} L_{\nu_{\gamma}})
\int u_{diff} \, dr \, .
\eqno(4) $$
The value of $L_{BC}$ can be easily calculated by assuming that 
$u_{diff} \simeq u_{BEL}$. At $r \le r_{BEL}$, where $r_{BEL}$
is the distance of the broad emission line region,
$u_{BEL} \simeq 3 \times 10^{-3}$erg cm$^{-3}$ (Peterson 1993) and 
drops very fast at $r>r_{BEL}$.
In this case, assuming $\theta_{obs} = 1/\Gamma$, Eq.(4) gives 
$$ L_{BC}^{(BEL)} \simeq 6.8 \times 10^{46} {\rm erg \, s}^{-1}
\, {r_{BEL} \over 3 \times 10^{17} {\rm cm} } \, {\Gamma \over 15} \,
{\nu_{\gamma} L_{\nu_{\gamma}} \over 10^{48} {\rm erg \, s}^{-1}}
\,  . \eqno(5) $$
This luminosity should peak at $h\nu_{BC} \sim 2 (\Gamma/15)^2 
(\bar \nu_{BEL}/10{\rm eV})$ keV;  its magnitude is already on the 
order of the observed soft X-ray luminosities in FSRQ. It implies 
that at distances $r < 10^{17}$cm, where bulk Comptonization of a 
direct disc radiation would strongly exceed $L_{BC}^{(BEL)}$, the jet 
is still not fully developed (accelerated/collimated) and/or the 
inner parts of a disc are radiatively inefficient, in turn suggesting 
that the outward transport of angular momentum occurs via other 
means (e.g. a disk wind) rather than viscosity in the disk.  
\smallskip

\noindent
{\it Discrete ejecta}
 
In the simplest version of the popular internal shock model for 
production of $\gamma$-ray flares, the dissipative events 
involve collisions of ejecta propagating down
the jet with different velocities. In this case, prior to the collision,
the cold ejecta produce soft X-ray flares with the apparent luminosity
$$ L_{BC,i} \simeq 
{\cal D}_i^4 \Gamma_i^2 \xi_i N_{e,i} c \sigma_T u_{diff}   \, , \eqno(6) $$
where $i =1,2$ denotes the ejecta assumed to move prior to the collision
with $\Gamma_2 > \Gamma_1 \gg 1$, ${\cal D}_i$ are the respective Doppler 
factors and $\xi_i$ is the fraction of $N_{e,i}$ contributing
to the radiation observed at a given instant (see Sikora \& Madejski 2002).
These soft X-ray flares are predicted to precede the non-thermal flares by
$$\delta t_i  \sim {r_{fl} \over c \Gamma_i {\cal D}_i} \sim
{{\cal D} \Gamma \over {\cal D}_i \Gamma_i} t_{fl}  \, . \eqno(7)$$

Assuming that the number of electrons is the same in both ejecta 
($N_{e,i} = N_e/2$), we obtain using Eqs. (2) and (6) 
$$ L_{BC,i}^{(BEL)} \simeq { {\cal D}_i^4 \Gamma_i^2 \Gamma^2 \over {\cal D}^5}
{c \sigma_T u_{BEL} \over m_e c^2} (\nu_{\gamma} L_{\nu_{\gamma}}) t_{fl} 
\xi_i \, , \eqno(8) $$ 
where, for ejecta with equal masses and proper lengths, $\lambda_0$,
$\xi_i \simeq {\rm Min}[1, r_{BEL}/(\lambda_0 {\cal D}_i)]$.
From the shock model, $\lambda_0= g_0 c {\cal D} t_{fl} $, where
$g_0$ depends on $\Gamma_2/\Gamma_1$ and on the adiabatic index 
$\hat \gamma$ of the shocked plasma. For  $\Gamma_2/\Gamma_1 = 2.5$ and 
$\hat \gamma = 5/3$, $g_0 \simeq 0.64$. With these particular parameters 
and noting that 
$\Gamma \simeq \sqrt{\Gamma_2\Gamma_1}$, one can find that for
$\Gamma=15$ and  $\theta_{obs}=1/\Gamma$ the precursors would have 
luminosities 
$ L_{BC,1}^{(BEL)} \sim 7.6 \times 10^{45}$ erg s$^{-1}$ and 
$L_{BC,12}^{(BEL)} \sim 4.7 \times 10^{46}$ erg s$^{-1}$, would 
precede the non-thermal flares by $\delta t_1 \sim 1.7 t_{fl}$ 
and $\delta t_2 \sim 0.7 t_{fl}$, and their spectra would peak 
at $\nu_{BC,1} \simeq 1.3$ keV and $\nu_{BC,2} \simeq 3.2$ keV.   

Additional contribution to $L_{BC,i}$ from Comptonization of  
direct radiation of the accretion disc would lead to precursors 
so prominent that they should have been detected in the available data.  
The lack of any such detections implies that, just as in the 
steady state case, the acceleration phase extends up to distances 
$r > 10^{17}$ cm, and/or that central parts of the disc are radiatively
inefficient.  
Such precursors should be easily detected by the current missions 
such as XMM even if they are produced at $r>10^{17}$cm, 
unless the  nonthermal flares arise from instabilities 
triggered {\sl in situ}, rather than due to modulation of the flow 
by the central engine.

It should be emphasized here that all above results do not {\sl explicitly} 
depend on the proton number and magnetic field intensities and 
on  related issues as the pair content and the jet power.  These 
aspects become crucial when the mechanisms and the energetics of 
the dissipative events are addressed. 
 
\acknowledgements
M.S. thanks to SOC for their invitation and generous hospitality. 
Support under the Chandra NASA grant (SAO number GO1 2113X) and 
Polish KBN grant 5 P03D 002 21 is gratefully acknowledged.

\end{document}